\documentclass[longauth]{aa}  

\usepackage{graphicx}
\usepackage{txfonts}
\usepackage{graphicx}
\usepackage{txfonts}
\usepackage{graphicx}
\usepackage{lscape}
\usepackage{longtable}
\usepackage{natbib}
\usepackage{color}
\usepackage{array} 
\usepackage{array} 
\usepackage{tikz,array}
\usetikzlibrary{calc}
\usepackage{txfonts}
\usepackage{color}
\usepackage{multirow}
\usepackage{here}

\usepackage{txfonts}
\usepackage{amsmath,amstext}
\usepackage{graphicx}
\usepackage{epstopdf}
\usepackage{float}
\usepackage{array}
\usepackage{mathtools}
\usepackage{booktabs}
\usepackage{subfigure}
\usepackage{url}
\usepackage{helvet}
\usepackage{tabularx}
\usepackage{multirow}
\usepackage{natbib}
\usepackage[flushleft]{threeparttable}
\usepackage{lscape}
\usepackage{pdflscape}
\usepackage{longtable}
\usepackage{wasysym}
\usepackage{float}
\usepackage{relsize}
\usepackage{color}
\usepackage{breqn}
\usepackage{bm}
\usepackage{enumitem}

\usepackage[colorlinks, citecolor=blue, linkcolor=blue]{hyperref}
\hypersetup{colorlinks,breaklinks, linkcolor=blue,urlcolor=magenta, anchorcolor=blue,citecolor=blue}

\def\ms{\hbox{m s$^{-1}$}}         
\def\cms{\hbox{cm\;s$^{-1}$}}       
\def\gcm3{\hbox{g cm$^{-3}$}}       
\def\Msun{\hbox{$\mathrm{M}_{\astrosun}$}}             
\def\Rsun{\hbox{$\mathrm{R}_{\astrosun}$}}
\def\Mjup{\hbox{$\mathrm{M}_{\rm Jup}$}}

\def\Mearth{\hbox{$\mathrm{M}_{\oplus}$}}
\def\Rearth{\hbox{$\mathrm{R}_{\oplus}$}}

\bibpunct{(}{)}{;}{a}{}{,} 

\newcommand{\be}{\begin{equation}}
\newcommand{\ee}{\end{equation}}

\newcommand{\hd}{HD\,22496}
\newcommand{\hdb}{HD\,22496\,b}

\usepackage{romannum}

\begin{document}

   \title{HD\,22496\,b: the first ESPRESSO standalone planet discovery\thanks{Based on Guaranteed Time Observations collected at the European Southern Observatory (ESO) under ESO programs 1102.C-074, 1104.C-0350, and 106.21M2 by the ESPRESSO Consortium.}}


   \author{
J.~Lillo-Box\inst{\ref{cab}}, 
J.~P. Faria\inst{\ref{depfisporto},\ref{porto}},
A.~Su\'arez Mascare\~no\inst{\ref{iac},\ref{ull}}, 
P.~Figueira\inst{\ref{eso},\ref{porto}},                
S.G.~Sousa\inst{\ref{porto}},
H.~Tabernero\inst{\ref{cab-torrejon}},     
C.~Lovis\inst{\ref{geneva}},
A.M.~Silva\inst{\ref{depfisporto},\ref{porto}},
O.D.S.~Demangeon\inst{\ref{depfisporto},\ref{porto}} 
S.~Benatti\inst{\ref{inaf-palermo}},
N.C.~Santos\inst{\ref{depfisporto},\ref{porto}}
A.~Mehner\inst{\ref{eso}},                
F.A.~Pepe\inst{\ref{geneva}},
A.~Sozzetti\inst{\ref{inaf-torino}}, 
M.R.~Zapatero~Osorio\inst{\ref{cab-torrejon}},
J.I.~Gonz\'alez Hern\'andez\inst{\ref{iac},\ref{ull}},
G.~Micela\inst{\ref{inaf-palermo}},
S.~Hojjatpanah\inst{\ref{porto},\ref{marseille}},
R.~Rebolo\inst{\ref{iac},\ref{ull},\ref{csic}}, \\
S.~Cristiani\inst{\ref{inaf-trieste}},
V.~Adibekyan\inst{\ref{depfisporto},\ref{porto}},
R.~Allart\inst{\ref{montreal},\ref{geneva}},
C.~Allende Prieto\inst{\ref{iac},\ref{ull}},
A.~Cabral\inst{\ref{ia-lisboa},\ref{depfis-lisboa}}
M.~Damasso\inst{\ref{inaf-torino}}, \\
P.~Di~Marcantonio\inst{\ref{inaf-trieste}}, 
G.~Lo~Curto\inst{\ref{eso}},
C.J.A.P.~Martins\inst{\ref{porto},\ref{caup}},
D.~Megevand\inst{\ref{geneva}}, 
P.~Molaro\inst{\ref{inaf-trieste},\ref{trieste-physics}},\\
N.J.~Nunes\inst{\ref{ia-lisboa}},
E.~Pall\'e\inst{\ref{iac},\ref{ull}}, 
L.~Pasquini\inst{\ref{eso-garching}},
E.~Poretti\inst{\ref{tng},\ref{inaf-merate}},
S.~Udry\inst{\ref{geneva}}
}

\institute{
Centro de Astrobiolog\'ia (CAB, CSIC-INTA), Depto. de Astrof\'isica, ESAC campus, 28692, Villanueva de la Ca\~nada (Madrid), Spain\label{cab} \email{Jorge.Lillo@cab.inta-csic.es  }
\and Depto. de F\'isica e Astronomia, Faculdade de Ci\^encias, Universidade do Porto, Rua do Campo Alegre, 4169-007 Porto, Portugal \label{depfisporto}
\and Instituto de Astrof\' isica e Ci\^encias do Espa\c{c}o, Universidade do Porto, CAUP, Rua das Estrelas, PT4150-762 Porto, Portugal \label{porto} 
\and Instituto de Astrof\'{\i}sica de Canarias (IAC), Calle V\'{\i}a L\'actea s/n, E-38205 La Laguna, Tenerife, Spain \label{iac}
\and Departamento de Astrof\'{\i}sica, Universidad de La Laguna (ULL), E-38206 La Laguna, Tenerife, Spain \label{ull}
\and European Southern Observatory, Alonso de Cordova 3107, Vitacura, Region Metropolitana, Chile \label{eso}
\and Centro de Astrobiolog\'\i a (CSIC-INTA), Crta. Ajalvir km 4, E-28850 Torrej\'on de Ardoz, Madrid, Spain \label{cab-torrejon}
\and D\'epartement d'astronomie de l'Universit\'{e} de Gen\`{e}ve, Chemin Pegasi 51, 1290 Versoix, Switzerland \label{geneva}
\and INAF - Osservatorio Astronomico di Palermo, Piazza del Parlamento 1, I-90134 Palermo, Italy\label{inaf-palermo}
\and INAF - Osservatorio Astrofisico di Torino, via Osservatorio 20, 10025 Pino Torinese, Italy \label{inaf-torino}
\and Aix Marseille Univ, CNRS, CNES, LAM, Marseille, France \label{marseille}
\and Consejo Superior de Investigaciones Cient\'{\i}cas, Spain \label{csic}
\and INAF - Osservatorio Astronomico di Trieste, via G. B. Tiepolo 11, I-34143 Trieste, Italy \label{inaf-trieste}
\and Department of Physics, and Institute for Research on Exoplanets, Universit\'e de Montr\'eal, Montr\'eal, H3T 1J4, Canada \label{montreal}
\and  Instituto de Astrof\' isica e Ci\^encias do Espa\c{c}o, Faculdade de Ci\^encias da Universidade de Lisboa, Campo Grande, 1749-016, Lisboa, Portugal \label{ia-lisboa}
\and Faculdade de Ci\^encias da Universidade de Lisboa (Departamento de F\'isica), Edif\'icio C8, 1749-016 Lisboa, Portugal\label{depfis-lisboa}
\and Centro de Astrof\'{\i}sica da Universidade do Porto, Rua das Estrelas, 4150-762 Porto, Portugal \label{caup}
\and Institute for Fundamental Physics of the Universe, Via Beirut 2, I-34151 Miramare, Trieste, Italy \label{trieste-physics}
\and ESO, Karl Schwarzschild Strasse 2, 85748 , Garching bei Muenchen, Germany \label{eso-garching}
\and Fundaci\'on G. Galilei -- INAF (Telescopio Nazionale Galileo), Rambla J. A. Fern\'andez P\'erez 7, E-38712 Bre\~na Baja, La Palma, Spain \label{tng}
\and INAF - Osservatorio Astronomico di Brera, Via E. Bianchi 46, I-23807 Merate, Italy \label{inaf-merate}
}

   \date{Received XXX; accepted YYY}

 
  \abstract
   {The ESPRESSO spectrograph is a new powerful tool to detect and characterize extrasolar planets. Its design allows to reach an unprecedented radial velocity precision (down to a few tens of \cms{}) and long-term thermo-mechanical stability.}
   {We present the first standalone detection of an extrasolar planet by blind radial velocity search using ESPRESSO {and aim at showing the power of the instrument in characterizing planetary signals at different periodicities in long observing time spans.}}
   {We use 41 ESPRESSO measurements of \hd{} obtained {within a time span of 895~days} with a median photon noise of 18~\cms{}. A radial velocity analysis is performed to test the presence of planets in the system and to account for the stellar activity of this K5-K7 main sequence star. {For benchmarking and comparison, we attempt the detection with 43 archive HARPS measurements and compare the results yielded by the two datasets.} We also use four TESS sectors to search for transits.}
   {We find radial velocity variations compatible with a close-in planet with an orbital period of $P=5.09071\pm0.00026$~days when simultaneously accounting for the effects of stellar activity at longer time scales ($P_{\rm rot}=34.99^{+0.58}_{-0.53}$~days). We characterize the physical and orbital properties of the planet and find a minimum mass of $5.57^{+0.73}_{-0.68}$~\Mearth{}, right in the dichotomic regime between rocky and gaseous planets. {Although not transiting according to TESS data}, if aligned with the stellar spin axis, the absolute mass of the planet must be below 16~\Mearth{}. We find no significant evidence for additional signals in the data with semi-amplitudes {above 56~\cms{} at 95\% confidence.}}
   {With a modest set of radial velocity measurements, ESPRESSO is capable of detecting and characterizing low-mass planets and constrain the presence of planets in the habitable zone of K-dwarfs down to the rocky-mass regime.}


	\titlerunning{HD\,22496: the first ESPRESSO planet}
	\authorrunning{Lillo-Box et al.}

   \maketitle
%

\section{Introduction}

The field of extrasolar planet detection through the radial velocity (hereafter RV) technique has evolved very fast during the past decades. This evolution has been driven by technical developments, especially in the wavelength calibration sources (e.g., \citealt{cersullo19,coffinet19}) and the instrumental stability (e.g., \citealt{mayor03}). 
First the iodine-cell, then the ThAr and now the Fabry-Perot and laser frequency comb technologies are helping us to reach the few tens of \cms{} precision in RV. The final precision of the instrument is also determined by the stability of the ambient conditions. Great efforts have been made in this regard in the past years, using vacuum chambers to isolate the instrument from external environmental changes (e.g., HARPS - \citealt{mayor03}, CARMENES - \citealt{quirrenbach10}, or HIRES - \citealt{vogt94}). The state-of-the-art of this evolution is the ESPRESSO instrument \citep{pepe20} at the Paranal Observatory. This instrument collects the light from any (or all) the Unit Telescopes from the Very Large Telescope (VLT) through the coud\'e trains \citep{cabral10}, which feed the instrument together with simultaneous wavelength calibration sources. The instrument is located in a three-layer isolation room where the ambient conditions (pressure, temperature and humidity) are kept stable \citep{alvarez18}. This allows ESPRESSO to reach a RV precision at the level of 10~\cms{} on sky (see \citealt{pepe20}).

The instrument has been available to the community from ESO's period P104 (October 2018). Since then, by using the {guaranteed} time observations (GTO), the ESPRESSO Consortium has shown the capability of the instrument {to characterize} the atmosphere of hot and warm Jupiters (e.g., \citealt{ehrenreich20,allart20,borsa21,santos20,tabernero21,casasayas-barris21}), determine precise masses of transiting planetary systems (e.g., \citealt{toledo-padron20,sozzetti21,mortier20}) and to deeply investigate the complex architecture of planetary systems like TOI-178 \citep{leleu19,leleu21}, {$\pi$ Mensae} \citep{damasso20} or Proxima Centauri \citep{suarez-mascareno20}. The community has also made use of this instrument for these purposes, with special mention to the largest RV dataset taken so far with ESPRESSO, 113 spectra of LHS\,1140 that allowed the estimation of the water mass fraction of its habitable zone planet (\citealt{lillo-box20b}). 

In this paper, we report the first {discovery} and characterization of a planet made by this instrument in the context of the GTO observations. In Sect.~\ref{sec:observations} we present the observations used. In Sect.~\ref{sec:Analysis} we describe the stellar and planet characterization methods, and the results are presented in Sect.~\ref{sec:Results}. We conclude in Sect.~\ref{sec:Conclusions}.

\section{Observations and general properties}
\label{sec:observations}

We observed \hd{} with ESPRESSO \citep{pepe20} within the context of the GTO granted to the ESPRESSO Consortium. The selection of this and other GTO targets surveyed in the context of blind search for planets is described in \cite{hojjatpanah19}. A total of 41 spectra with the HR21 mode and an exposure time of  900\,s each were obtained from Paranal observatory (ESO, Chile), {with a mean signal-to-noise ratio (S/N) of 270 at 550\,nm and simultaneously illuminating the second fiber with the Fabry-P\'erot}. The observations span 895 days between 24-oct-2018 and 06-apr-2021  and a typical cadence of one spectrum every 1-3 days for each campaign. We note that observations suffered from the long gap between March and December 2020 due to the pandemic situation of COVID-19. The first four data points were obtained prior to the {ESPRESSO fiberlink exchange in June 2019 that improved the efficiency of the instrument by more than 50\% as described in \cite{pepe20}. This intervention introduced an RV offset that leads to treating this dataset separately (labelled as ESPR18 in this paper). Besides, during the ramp-up after the pandemic closure of the observatory, a calibration lamp exchange on 17-dec-2020 introduced an additional offset in the RV measurements. Hence we consider the 16 spectra between June 2019 and December 2020 (labelled as ESPR19 along this paper) and the 21 observations obtained afterwards (ESPR21) as independent datasets.} We assume these three datasets as coming from different instruments to allow for possible RV offsets and different levels of RV jitter.

\begin{figure*}
\centering
\includegraphics[width=1\textwidth]{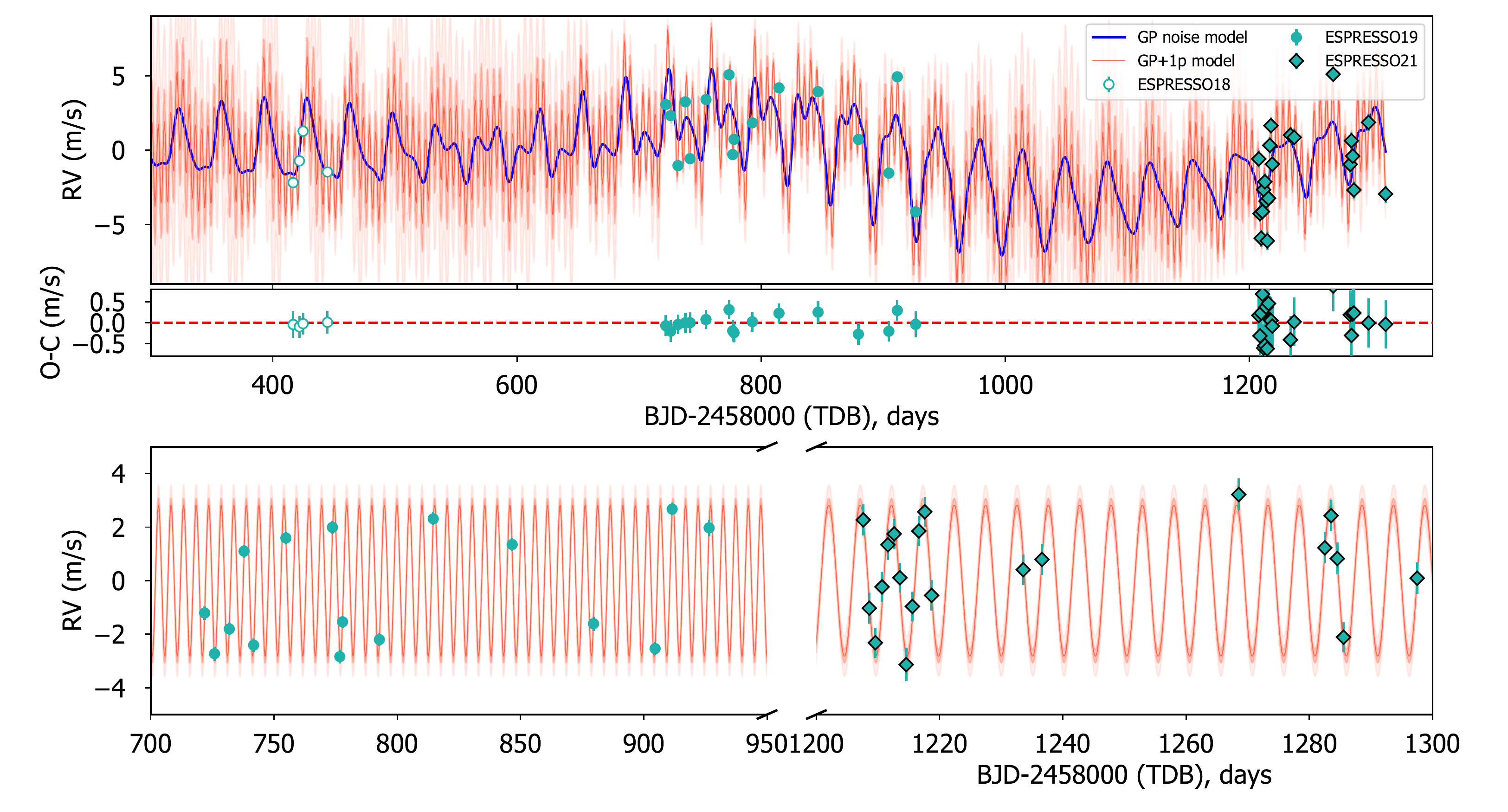}
\caption{ESPRESSO radial velocity time series for \hd{}. The upper panel shows the complete series with the three datasets corresponding to the three ESPRESSO windows (ESPR18, ESPR19 and ESPR21). The median GP model is shown in blue and the median GP+Keplerian model is shown in red, with 95\% confidence intervales shown as red shaded regions. The lower panel shows a close view to the ESPR19 (left) and ESPR21 (right) datasets after removing the median GP model. }
\label{fig:RVespr}
\end{figure*}

The data were reduced with the instrument Data Reduction Software (DRS) pipeline version 2.2.8\footnote{The ESPRESSO DRS is publicly available at \url{https://www.eso.org/sci/software/pipelines/index.html}} \citep{pepe20}. The pipeline also extracts the RV by using the cross-correlation technique \citep{baranne96} against a K5 mask as selected in this case. It also computes different activity indicators for each epoch, as described in \cite{pepe20}. A summary of the derived RV, FWHM and BIS is shown in Table~\ref{tab:RVespresso}. The median RV uncertainty of this dataset corresponds to 18 \cms{} (and a mode of 15 \cms{}) and we found a scatter in the dataset corresponding to 2.7 \ms{} (more than one order of magnitude larger than the median uncertainty), a first indication of an additional source of variability. The ESPRESSO RV time series of \hd{} is shown in Fig.~\ref{fig:RVespr}. 

\hd{} was also observed by HARPS \citep{mayor03} in the past, on lower-precision RV campaigns aiming at a multitude of objectives (program IDs 072.C-0488 -- PI: M. Mayor--, 085.C-0019 -- PI: G. Lo Curto--, and 183.C-0972 -- PI: S. Udry). In total, 43 HARPS spectra are publicly available from the ESO archive. The radial velocities and activity indicators were computed by the instrument DRS v3.5. Given the relatively short exposure times (ranging from 90\,s to 900\,s with a median of 120\,s), the median uncertainty of this dataset corresponds to 2.2 m/s and a standard deviation of 4.9 m/s (see in Table~\ref{tab:RVespresso}).

TESS \citep{ricker14} observed \hd{} during sectors 3, 4, 30, and 31. We retrieved the light curves from these sectors through the MAST archive\footnote{\url{https://mast.stsci.edu/portal/Mashup/Clients/Mast/Portal.html}} and used the detrended Pre-search Data Conditioned Simple Aperture Photometry (PDCSAP) flux provided by the TESS SPOC pipeline \citep{jenkins16}. {ESPRESSO18 observations were obtained during TESS Sector 4, ESPRESSO19 around seven months before Sector 30, and ESPRESSO 21 observations started around 3 months after Sector 31. Additional description of the light curve extraction and analysis is provided in Appendix~\ref{app:TESS}.}

\section{Analysis}
\label{sec:Analysis}

\subsection{Stellar properties}
\label{sec:stellarprop}

\hd{} (also known as LHS\,1563, GJ\,146, or HIP\,16711) is a bright ($V=8.9$~mag) late-type K-dwarf star in the solar vicinity. The \textit{Gaia} \citep{gaia} EDR3 data release \citep{gaia21} provides a precise parallax of $\pi=73.520\pm0.016$~mas \citep{lindegren20}, corresponding to a distance of $d=13.602\pm0.003$~pc {(see Table~\ref{tab:basic}). According to the \textit{Gaia} proper motions and the relations from \cite{bensby03}, this star likely belongs to the galactic thin disk (with a probability 67 times higher than belonging to the thick disk)}. The photometric information from the second data release (DR2, \citealt{gaia18}) also provides an effective temperature of 4250~K and $\log{g}=4.5$~dex, corresponding to a K5-K7 main-sequence star. \cite{mann15} derived a mass and radius of M$_{\star}=0.684$~\Msun{} and R$_{\star}=0.45$~\Rsun{} through spectroscopic characterization. Based on the comparison between \textit{Gaia} and \textit{Hipparcos} proper motion differences, \cite{kervella19} put constraints on potential companions to this star, and set {a sensitivity of $0.45\pm0.25$~\Mjup{} at 1 au}.

We combined all ESPRESSO observations of \hd{} into a single high S/N spectrum to estimate the stellar atmospheric parameters (namely effective temperature - $T_{\rm eff}$-, surface gravity - $\log{g}$-,  and metallicity - [Fe/H]). We used the spectral synthesis method by means of the {\scshape SteParSyn} code \citep{tab18,tab21a}. We employed a grid of synthetic spectra computed with the {\tt Turbospectrum} \citep{ple12} code alongside MARCS stellar atmospheric models \citep{gus08} and atomic and molecular data of the Gaia-ESO line list \citep{hei21}. We employed a selection of $\ion{Fe}{i}$ lines following the line list given by \citet{tab19} for metal rich dwarf stars. {\sc SteParSyn} allowed us to compute the following stellar atmospheric parameters: $T_{\rm eff}$~$=$~4385~$\pm$~21~K, $\log{g}$~$=$~4.69~$\pm$~0.05~dex, [Fe/H]~$=$~$-$0.08~$\pm$~0.02~dex.  

{The values reported in the TESS Input Catalog (TIC v8.0.1, \citealt{stassun19}, T$_{\rm eff}\sim 4102$~K and $\log{g}\sim4.52$~dex), are not consistent with the ones derived in our spectroscopic analysis. However our spectroscopic analysis is based on a very high resolution, and high signal-to-noise ESPRESSO spectrum, compared to the photometric characterization in the TIC catalog. Moreover, using Gaia EDR3 paralaxes and photometry we obtain a trigonometric surface gravity  of $4.73\pm0.02$~dex, consistent with the value derived by our spectrosocopic analysis.}

{We used the TESS light curve to explore the level of stellar variability and estimate a stellar rotation period. In Appendix~\ref{app:TESS} we provide a detailed description of the analysis performed. However, this in depth study is still inconclusive, although it points to a stellar rotation period in the range $P_{\rm rot} = 30^{+38}_{-18}$ days.}

\subsection{Radial velocity analysis}
\label{sec:RVanalysis}

\begin{figure}
\centering
\includegraphics[width=0.5\textwidth]{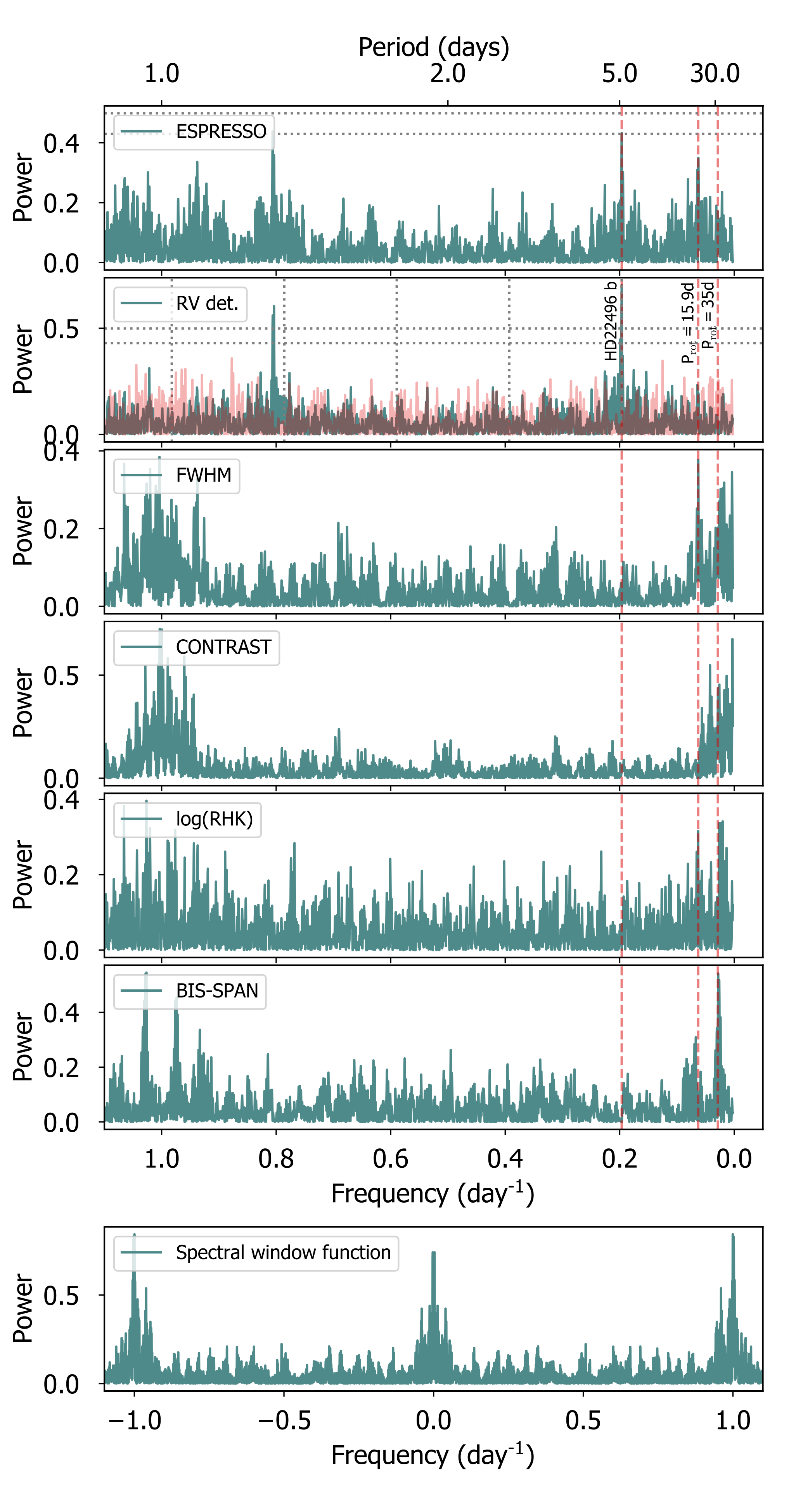}
\caption{Periodogram of ESPRESSO time series. From top to bottom: ESPRESSO RV, ESPRESSO RVs aftera simple weighted linear detrending with the BIS-SPAN activity indicator (green color) and after removing a Keplerian signal with $P=5.09$ days (red), FWHM of the CCF, CONTRAST of the CCF, log(RHK), BIS-SPAN and the window function. The dashed red vertical lines marks the location of the confirmed planet \hdb{}, the stellar rotation period found by the GP and its alias. In the two upper panels, the dotted horizontal lines mark the 0.1\% and 1\% false alarm probability levels. In the second panel, the vertical dotted lines mark the first 10 alias periods of the \hdb{} signal.}
\label{fig:periodogram}
\end{figure}

{The generalized lomb-scargle periodogram of the ESPRESSO RV time series is shown in Fig.~\ref{fig:periodogram} (upper panel). It shows a forest of periodicities around 30 days, similar to the expected rotation period from the $\log{R^{\prime}_{\rm HK}}$ relations from \cite{suarez-mascareno15} (35 days) and the same parameter space derived form the TESS light curve (see Appendix~\ref{app:TESS})}. Indeed, the periodogram of the different activity indicators (bisector span - BIS-, full-width at half-maximum - FWHM - of the CCF, the CCF contrast and the $\log{R^{\prime}_{\rm HK}}$) also show significant peaks {at 35 days and its potential 1-day aliases of 15.9 days and 70 days (see panels 3 to 6 of Fig.~\ref{fig:periodogram})}. The correlation of the CCF-FWHM with the RV time series yields a {Pearson's coefficient of 0.71, hence suggesting that part of the RV variations are due to stellar activity. The periodogram also reveals a clear signal at 5.09 days and its 1-day alias at 1.2 days. The power of both signals is clearly enhanced by performing a simple weighted linear detrending with the CCF-FWHM indicator (see second panel of Fig.~\ref{fig:periodogram}). }

Based on this preliminary analysis, we analyze the ESPRESSO RV data by assuming different possible scenarios with different number of planets and orbital configurations. In particular, we test four models, including no planets (0p, null hypothesis), one planet in circular orbit (1p1c), one planet in eccentric orbit (1p), and two planets in circular orbits (2p1c2c). 


We account for stellar activity by using a Gaussian Process (GP) approach where the CCF-FWHM activity indicator is used as a proxy\footnote{We have also tried the bisector span measured on the CCF as an activity proxy with similar results as the FWHM.} (\citealt{suarez-mascareno20}). 

The Keplerian signals are modeled by a period ($P_i$), time of inferior conjunction ($T_{0,i}$), and semi-amplitude ($K_i$) for each of the planets tested, and the eccentricity ($e_i$) and argument of the periastron ($\omega_i$) for non-circular models. Additionally, we use a RV offset ($\delta_{\rm RV,j}$) and a jitter term ($\sigma_{\rm RV,j}$) for each instrument considered (four in total, taking into account HARPS and the three ESPRESSO epochs). Since we are using the CCF-FWHM as a proxy, we also add an offset ($\delta_{\rm FWHM,j}$) and jitter ($\sigma_{\rm FWHM,j}$) per instrument and epoch for these data. 
We use a quasi-periodic kernel \citep{ambikasaran14,faria16} for both the RV and the FWHM, both sharing the rotation period ($\eta_3$), time scale of the variation ($\eta_2$), and the scaling factor ($\eta_4$); while each having its own amplitude ($\eta_{\rm 1,RV}$ and $\eta_{\rm 1,FWHM}$). 


{We use a broad Gaussian prior for the inner planet centered at $\sim$5-days and with a one day width, while for the outer component in the two-planet model we assume a uniform prior between 5 and 200 days. Nevertheless, we note that broad uninformative priors were also tested for the inner component (uniform between 1.1 and 20 days) and for the GP hyper-parameter associated to the stellar rotation period, $\eta_3$ (log-uniform between 8 and 100 days). The exploratory results on this two parameters show strong evidence for the 5.09 days Keplerian signal against the null hypothesis and against the $\sim$1.25-day alias, and for the $\eta_3\sim$35 days against other possible periodicities also present in the periodogram of the activity indicators (see Fig.~\ref{fig:periodogram} and discussion above).}

We use the \texttt{emcee} Markov-Chain Monte-Carlo (MCMC) affine invariant ensamble sampler \citep{emcee} to populate the posterior distribution of the different parameters involved. For each model, we use a first burn-in phase of 100\,000 steps and four times as many walkers as number of parameters. A second phase (the production phase) contains 50\,000 steps instead. This is enough to ensure the convergence of the chains. This is checked by estimating the autocorrelation time and the corresponding chain length, with the latest being at least 20 times longer than the autocorrelation time to consider convergence. We then use 15\% of the final flattened chain (typically composed of $10^5-10^6$ elements) to estimate the Bayesian evidence of each model ($\ln{\mathcal{Z}_i}$) and its corresponding uncertainty through the \texttt{perrakis} implementation\footnote{\url{https://github.com/exord/bayev}. A python implementation by R. Di\'iaz of the formalism explained in \cite{perrakis14}.} \citep{diaz16}. 

This procedure is first performed on the ESPRESSO dataset (labelled as E) and subsequently on the whole ESPRESSO and HARPS dataset (labelled as H+E).


\section{Results}
\label{sec:Results}

\begin{figure}
\centering
\includegraphics[width=0.5\textwidth]{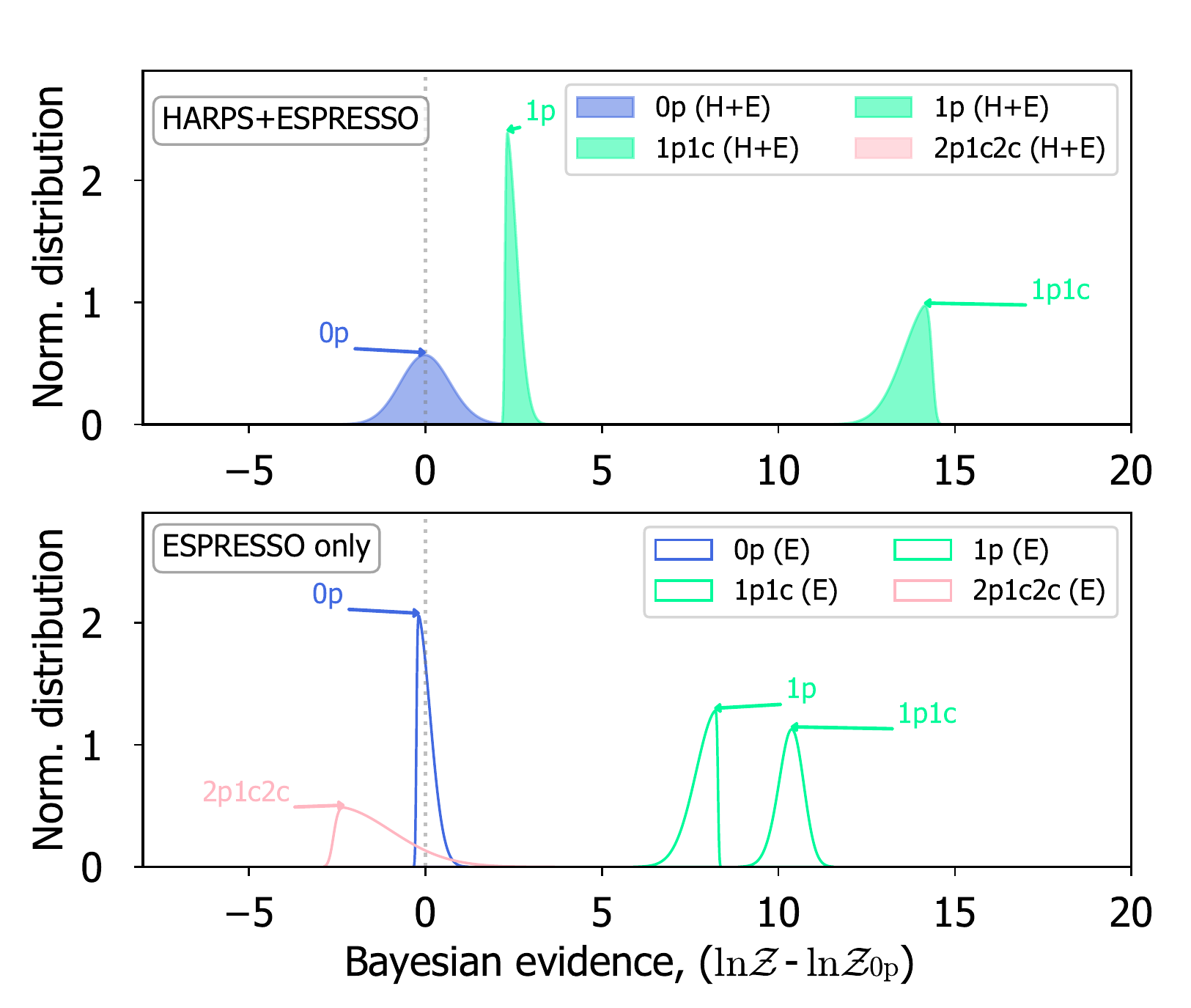}
\caption{Distribution of the log-Bayesian evidence for the different models tested using the Gaussian Process approach to account for the stellar activity (see Sect.\ref{sec:Analysis}). Filled shaded distributions correspond to the HARPS+ESPRESSO dataset (upper panel) while open distributions correspond to the ESPRESSO standalone dataset (lower panel). The median from the null hypothesis model for each dataset (labelled as 0p) has been subtracted. {In the upper panel, the two-planet model ("2p1c2c") has a very low odd ratio, hence lying outside of the plot limits.}}
\label{fig:evidence}
\end{figure}

\begin{figure}
\centering
\includegraphics[width=0.5\textwidth]{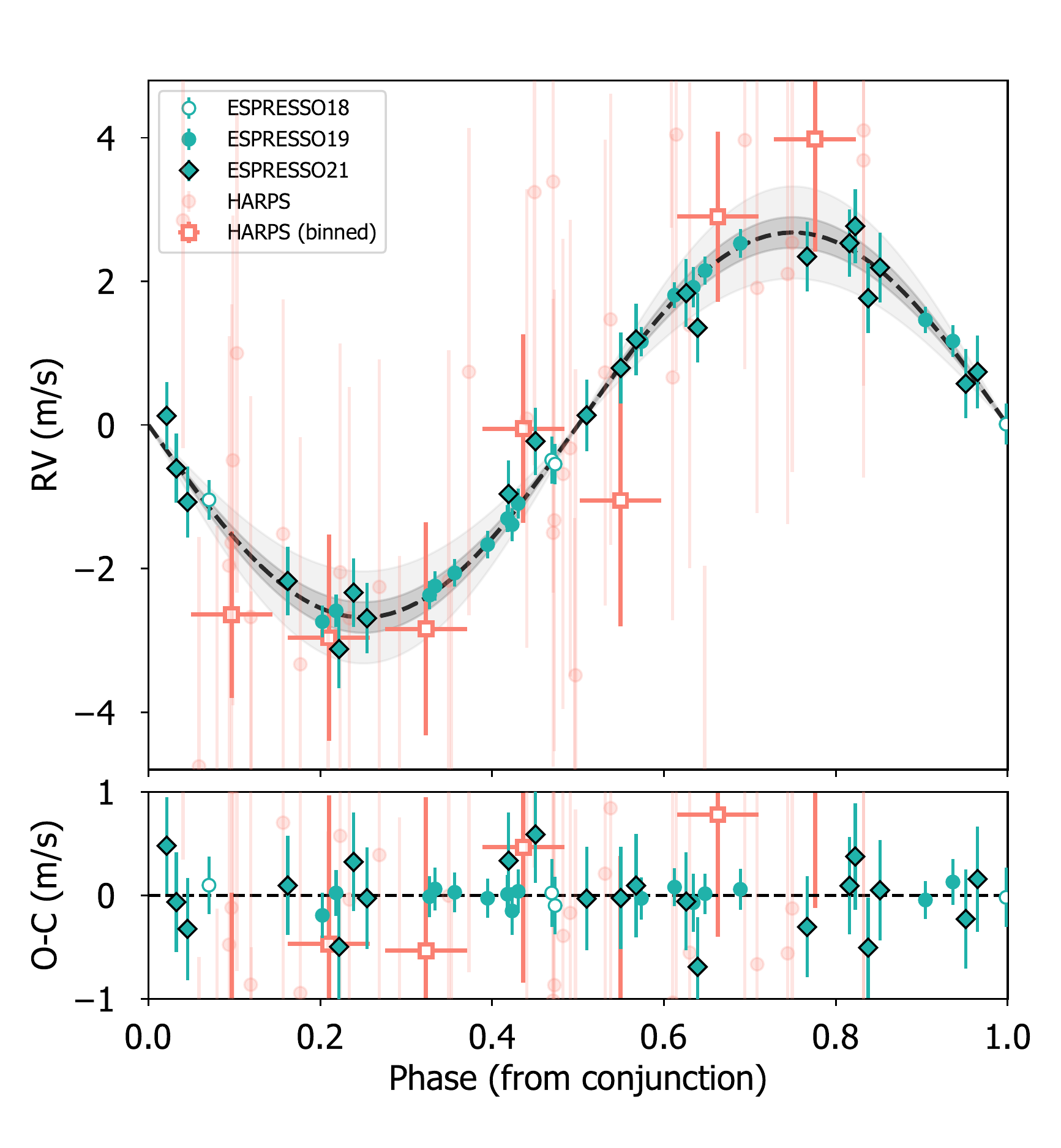}
\caption{Radial velocities from HARPS (red) and ESPRESSO (green) phase-folded with the period from planet \hdb{}. The red open symbols display represent binned values from HARPS measurements  corresponding to 10\% of the orbital phase. The median GP model has been subtracted. The bottom panel shows the residuals of the mean model.}
\label{fig:RVphase}
\end{figure}

 
The distribution of the Bayesian evidence values from the MCMC chains for the most relevant scenarios (those showing the largest values) is shown in Fig.~\ref{fig:evidence} relative to the null hypothesis (i.e., the model with no planets). {In this figure, we include the results for both the ESPRESSO and ESPRESSO+HARPS datasets.} As shown, the one-planet model assuming circular orbit (labelled as 1p1c) is strongly preferred against the null hypothesis ($\Delta\ln{\mathcal{Z}_{\rm 1p1c-0p}}>6$, \citealt{jeffreys98}) and against other more complex models. We note that the odds ratio between the 1p1c model and the null hypothesis surpasses the statistical significance standard threshold in both the ESPRESSO ($\Delta\ln{\mathcal{Z}_{\rm 1p1c-0p}}=+10.5$) and ESPRESSO+HARPS ($\Delta\ln{\mathcal{Z}_{\rm 1p1c-0p}}=+14.5$) datasets. We thus conclude that the one-planet circular orbit model is the one best supported by our datasets. {The ESPRESSO standalone dataset is enough to confirm this signal with statistical significance against the null hypothesis, and it was the one that originally allowed us to detect the planet signal. The addition of the HARPS data increases the significance of the detection, providing additional confidence that the planetary signal is present in the system by using an independent dataset. Consequently, we use the HARPS+ESPRESSO dataset for the subsequent discussion.} The phase-folded median model for this scenario is presented in Fig.~\ref{fig:RVphase} and the credible intervals for the different parameters involved are shown in Table~\ref{tab:posterior}. 

{The largest evidence model (the one-planet in circular orbit scenario with the ESPRESSO+HARPS dataset)} indicates that \hd{} is orbited by a planetary-mass object with a minimum mass of  $m_{\rm b}\sin{i_b} = 5.57^{+0.73}_{-0.68}$~\Mearth{}. Only an orbital inclination below $i<0.8^{\circ}$ (i.e., almost a fully face-on orbit) would place this companion above the planet-mass regime. Consequently, the probability of this being a planet signal (assuming random distribution for the orbital inclination) is 99.1\%. We can thus safely consider this companion to be of planetary nature. Additionally, we find that for orbital inclinations above $i>20^{\circ}$, the planet would have an absolute mass below the Neptune mass. On the other hand, based on the stellar rotation period and the upper limit of the projected stellar velocity provided in \cite{hojjatpanah19} (v$\sin{i}<2$~km/s), we can infer a lower limit for the inclination of the stellar spin of $i_{\star}>20^{\circ}$. Hence, {in the case of spin-orbit alignment}\footnote{{We note that this does not necessarily have to be the case, as demonstrated in many misaligned planetary systems including close-in components (e.g., \citealt{huber13b}).}}, the absolute mass of the planet will be in the range $m_{b} = 5.6-16$~\Mearth{}. This places the planet at the boundary between the super-Earth (rocky) and sub-Neptune (mostly gaseous) regimes.

The planet revolves around its star every 5.09 days and receives an insolation flux 22.8 times that of Earth. We tested the eccentric scenario and we found clear evidence in favor of the circular model{, with an odd ratio of $\Delta\ln{\mathcal{Z}_{\rm 1p1c-0p}}=+14.5$, as stated above}. Instead, we can set upper limits to the eccentricity of this planet up to $e_b<0.15$ {at the 95\% confidence level}. The GP catches the expected rotation period of the star, providing $\eta_3=34.99^{+0.58}_{-0.53}$~days, compatible with the expected value (see Sect.~\ref{sec:RVanalysis}). 

\begin{figure}
\centering
\includegraphics[width=0.5\textwidth]{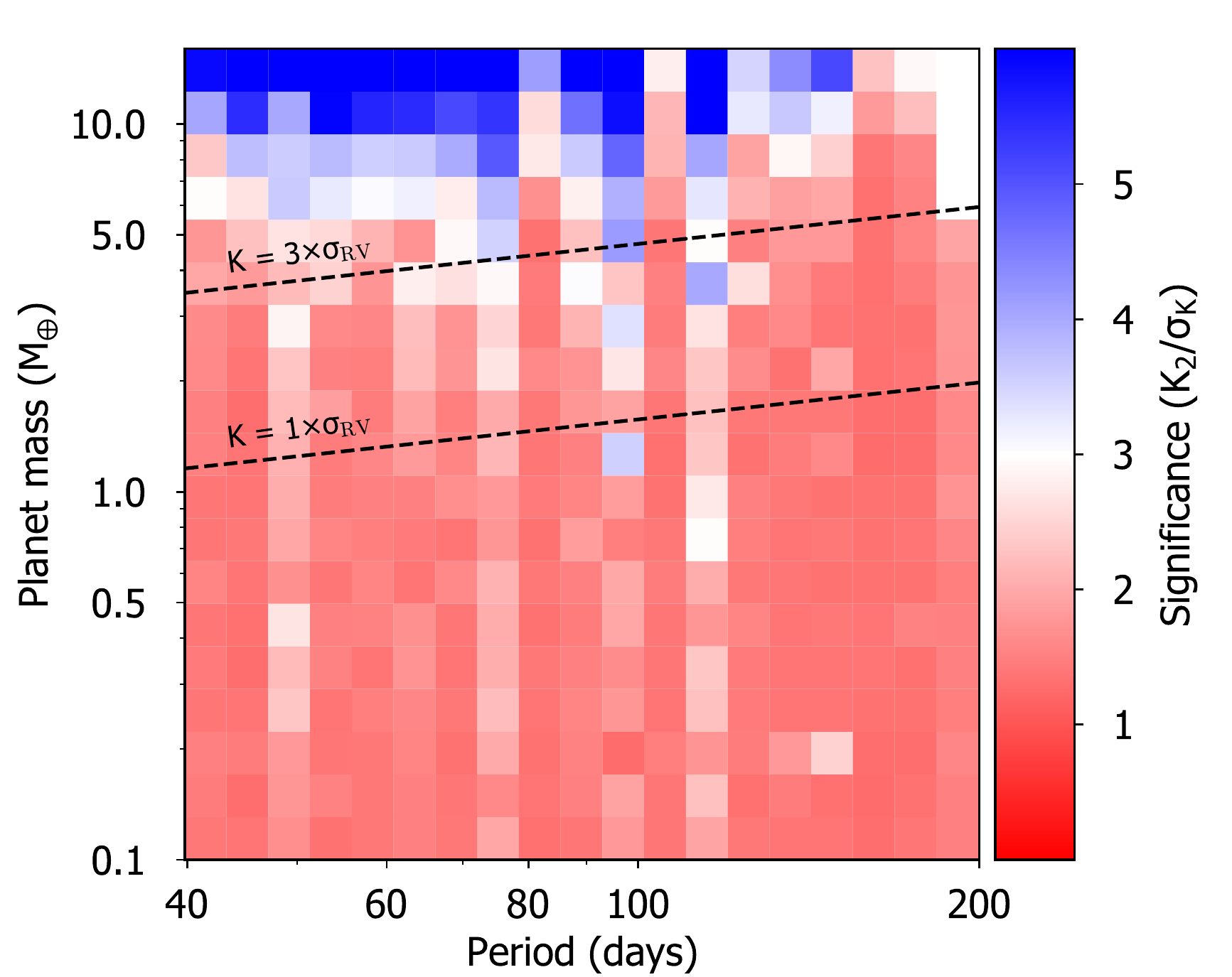}
\caption{Detectability matrix for the HARPS and ESPRESSO dataset within the habitable zone of \hd{} performed through injection-recovery. The color-code represents the significance in the posterior of the RV semi-amplitude parameter in units of its standard deviation, with ratios above 3 (signal detected) in blue. The dashed lines correspond to one and three times the standard deviation of the residuals from the one-planet circular model (i.e., 28 \cms{}).}
\label{fig:detectability}
\end{figure}

Additionally, we tested the detectability limits of planets inside the habitable zone of this star. To this end, we follow an injection-recovery analysis similar to that from \cite{lillo-box20b}. We simulate Keplerian signals at different periods (from 39 days to 200 days, within the habitable zone) and with different minimum masses (from 0.1~\Mearth{} to 16~\Mearth{}), assuming circular orbits. We add these signals to the ESPRESSO+HARPS dataset and explore their detectability by modeling the data in the same manner as explained in Sect.~\ref{sec:Analysis} {(i.e., including the GPs and the activity indicators as proxies for their hyper-parameters)}. Figure~\ref{fig:detectability} shows the resulting detectability matrix. In general terms, we can discard the presence of planets with minimum masses smaller than 5~\Mearth{} in the inner edge of the habitable zone (P$_{\rm HZ, out}\sim$40 days) and 7~\Mearth{} in the outer boundary (P$_{\rm HZ, out}\sim$200 days).

{Based on the above RV-driven analysis we searched for transits of the 5.09-day period planet (and any other potential signal) in the TESS light curve.} We used \texttt{wotan} \citep{wotan} to perform a smoothing of the photometric time series to remove long-term variations (using the \texttt{flatten} routine with a window length of 0.5-days) potentially due to either stellar activity or instrumental effects (especially apparent in sector 4). Then we searched for transit-like dimmings using the \texttt{tls} software \citep{heller19} on the detrended light curve. We found no significant peaks \texttt{tls} periodogram, suggesting that no transits are present in the data. We further phase-folded the TESS light curve by using the periodicity detected with the RV data (see Sect.~\ref{sec:RVanalysis}). No apparent transit is detectable above 50 parts per million depth. This implies a minimum orbital inclination for \hdb{} of $i<87.6^{\circ}$ or otherwise a maximum size of 0.34~\Rearth{}.

\section{Conclusions}
\label{sec:Conclusions}

We present the first standalone detection and confirmation of an extrasolar planet with the ESPRESSO instrument. With only 41 900\,s exposures, we reach a Bayesian evidence $>10^5$ times larger than the null model hypothesis ($\Delta\ln{\mathcal{Z}_{\rm 1p-0p}}=+14.5$), reaching a precision in the minimum mass of the planet of 13\%, below the typically required precision for atmospheric studies \citep{batalha19}. \hdb{} is {likely} a planet in the transition between super-Earth and sub-Neptune type planets ($m_{\rm b}\sin{i_b} = 5.57^{+0.73}_{-0.68}$~\Mearth{}) orbiting around a late K-type star. These stellar hosts are getting an increasing interest by the astrobiology community (see, e.g., the KOBE experiment\footnote{K-dwarfs Orbited By habitable Exoplanets (KOBE), PI: J. Lillo-Box, \url{https://kobe.caha.es/}}) as they represent friendly environments for life, with small stellar activity and low flare rate ($<1$\%, \citealt{gunther20b}), and the habitable zone sufficiently far from the star for the planets {to less likely be tidally locked \citep{barnes17}}, but sufficiently close to have relatively large probability of transit and be easy to detect through RV techniques. {The case of \hd{} however will only allow atmospheric studies through non-transiting techniques (e.g., phase curves with ARIEL, \citealt{ariel}) as the planet does not transit its host.}

With one close-in planet detected, \hd{} enters the group of late K-dwarfs with known planets. Indeed, planet occurrence rates from the Kepler mission, show that these K-dwarfs have 3-4 planets per star \citep{kunimoto20} and a high occurrence rate of 0.65 planets per star within the habitable zone, half of them being in the rocky regime. Hence, additional follow-up efforts should continue for this system to unveil the potential additional planets.

\begin{acknowledgements}
The authors acknowledge the ESPRESSO project team for its effort and dedication in building the ESPRESSO instrument. 
J.L-B. acknowledges financial support received from ”la Caixa” Foundation (ID 100010434) and from the European Union’s Horizon 2020 research and innovation programme under the Marie Skłodowska-Curie grant agreement No 847648, with fellowship code LCF/BQ/PI20/11760023. This research has also been partly funded by the Spanish State Research Agency (AEI) Projects No.ESP2017-87676-C5-1-R and No. MDM-2017-0737 Unidad de Excelencia "Mar\'ia de Maeztu"- Centro de Astrobiolog\'ia (INTA-CSIC).
J.P.F. is supported in the form of a work contract funded by national funds through FCT with reference DL57/2016/CP1364/CT0005. 
O.D.S.D. is supported in the form of work contract (DL 57/2016/CP1364/CT0004) funded by national funds through Fundação para a Ciência e Tecnologia (FCT).
A.M.S acknowledges support from the Fundação para a Ciência e a Tecnologia (FCT) through the Fellowship 2020.05387.BD. and POCH/FSE (EC).
ASM, CAP, JIGH, and RRL acknowledge financial support from the Spanish Ministry of Science and Innovation (MICINN) project AYA2017-86389-P.
This work was supported by FCT - Funda\c{c}\~ao para a Ci\^encia e a Tecnologia through national funds and by FEDER through COMPETE2020 - Programa Operacional Competitividade e Internacionaliza\c{c}\~ao by these grants: UID/FIS/04434/2019; UIDB/04434/2020; UIDP/04434/2020; PTDC/FIS-AST/32113/2017 \& POCI-01-0145-FEDER-032113; PTDC/FIS-AST/28953/2017 \& POCI-01-0145-FEDER-028953; PTDC/FIS-AST/28987/2017 \& POCI-01-0145-FEDER-028987. 
FPE and CLO would like to acknowledge the Swiss National Science Foundation (SNSF) for supporting research with ESPRESSO through the SNSF grants nr. 140649, 152721, 166227 and 184618. The ESPRESSO Instrument Project was partially funded through SNSF's FLARE Programme for large infrastructures.
The INAF authors acknowledge financial support of the Italian Ministry of Education, University, and Research with PRIN 201278X4FL and the "Progetti Premiali" funding scheme.
NJN acknowledges support form the following projects UIDB/04434/2020 \& UIDP/04434/2020, CERN/FIS-PAR/0037/2019, PTDC/FIS-OUT/29048/2017, COMPETE2020: POCI-01-0145-FEDER-028987, and FCT: PTDC/FIS-AST/28987/2017.

\end{acknowledgements}

%
%

\bibliographystyle{aa} 
\bibliography{../../../biblio2} 

\appendix

\section{TESS light curve analysis}
\label{app:TESS}
\begin{figure}
\centering
\includegraphics[width=0.5\textwidth]{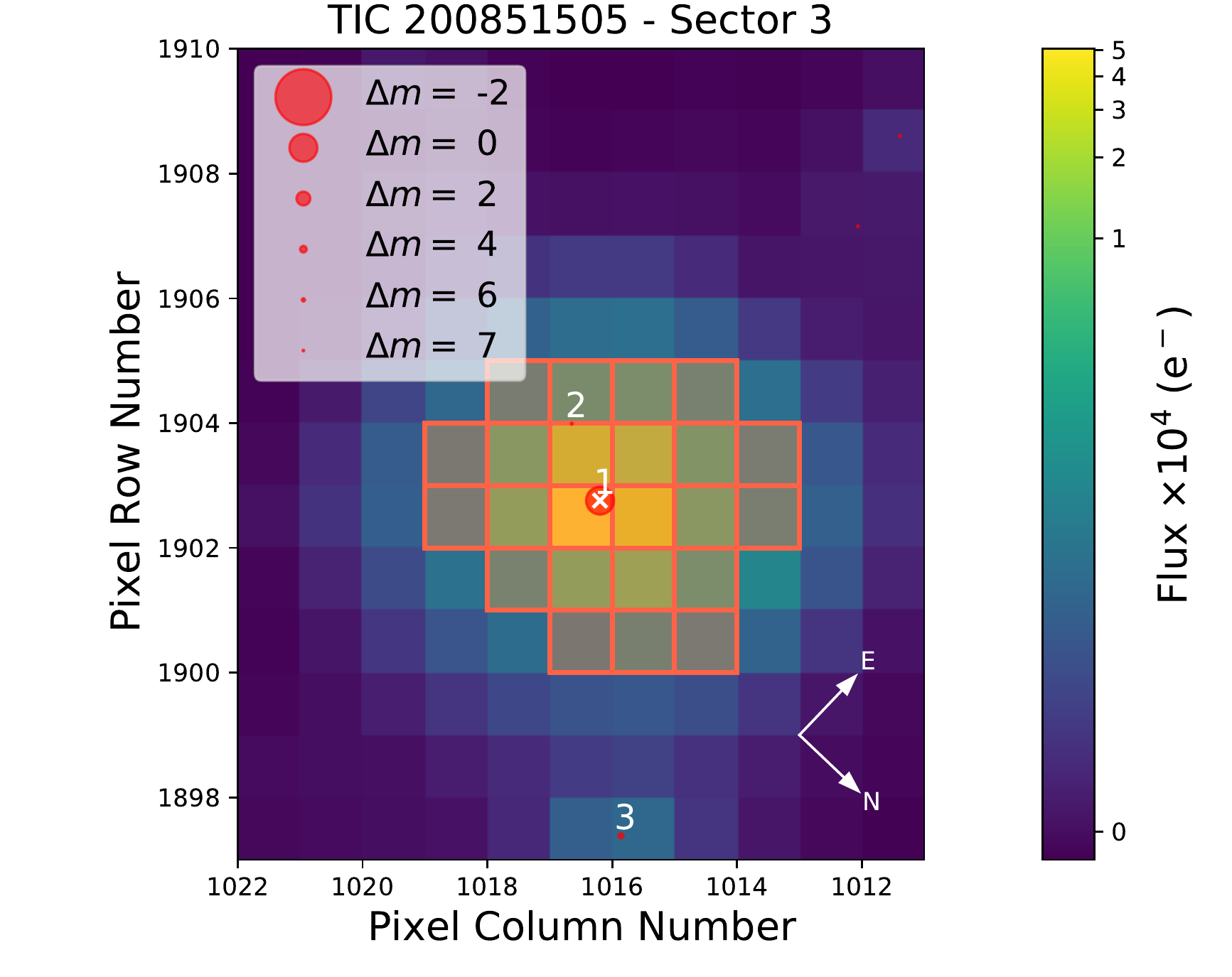}
\caption{Upper panel: Target Pixel File plot (using \texttt{tpfplotter} \citealt{aller20}) of \hd{} in Sector 3. The TESS pipeline aperture is marked with red shaded squares. The location of the target is marked by using a white cross and all additional sources identified by Gaia DR2 are marked with red circles (with sizes inversely proportional to their magnitude).} 
\label{fig:TESStpf}
\end{figure}

{TESS observations from sectors 3, 4, 30, and 31 were retrieved from the MAST archive and the PDCSAP detrended photometry was obtained.} We used \texttt{tpfplotter}\footnote{\url{https://github.com/jlillo/tpfplotter}} \citep{aller20} to first check for possible contamination within the TESS aperture and found only one source inside the aperture with contrast $\Delta G=7.4$~mag in the Gaia passband (see Fig.~\ref{fig:TESStpf}). This large contrast ensures a negligible contamination.

\begin{figure}
\centering
\includegraphics[width=0.5\textwidth]{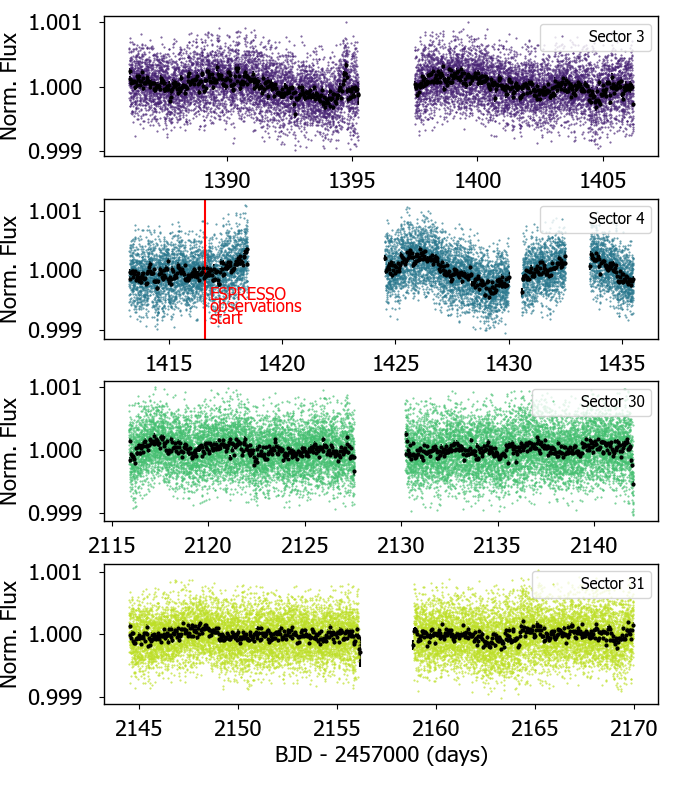}
\caption{{TESS light curve fro sectors 3, 4, 30, and 31 (from top to bottom). One-hour photometric bins are shown with black symbols. ESPRESSO observations of this target started on the date marked by the red vertical line (Sector 5) and ended two months after Sector 31.}}
\label{fig:TESS}
\end{figure}

{We excluded all exposures whose quality flags displays the bits 1, 2, 3, 4, 5, 6, 8, 10 and 12 as suggested by the TESS team. Based on the release notes of the TESS data\footnote{The release notes of all TESS sector including sectors 3, 4, 30 and 31 are available at \url{https://archive.stsci.edu/tess/tess_drn.html}.}, we also excluded all measurements of sector 4 taken before 1413.26 TJD because of a wrong pointing due to a erroneous guiding star table uploaded at the beginning of the sector. An interruption of communication between the instrument and the spacecraft also occurred towards the end of the first half of sector 4. The data collected after this interruption and before interruption at mid-sector for data downlink shows variations far greater than any other portions of LC in the SAP LC. The PDCSAP LC shows much smaller variations during this period, yet it still displays clear variations correlated with the signal observed in the SAP LC and which have shorter timescales than the rest of the LC. We thus excluded data taken between 1421 and 1423 TJD. Still in sector 4, we observe clear box-shape signals before the last three momentum dumps of the reaction wheels. The correlation between the momentum dumps and these events associated with the fact that such signals are not observed in the rest of the LC indicates that these signals are not of astrophysical origin. Consequently, we excluded all data taken between 1430 and 1430.5883 TJD, between 1432.5 and 1433.5883, and after 1435.5 TJD in sector 4. During sector 3, we observed in the PDCSAP LC an exponential decrease of the flux at the beginning of the sector and an increasing ramp-like signal after the data downlink interruption that are absent from the SAP LC. We suspect that these signals are due to an overcorrection and remove all measurements taken before 1386.113 TJD in sector 3 and between 1396.50 and 1397.50 TJD. The resulting light curve is shown in Fig.~\ref{fig:TESS}.}

{We then inspected the normalized TESS LC for signs of stellar activity induced modulation. We used the start and the end of each sector and the time of each momentum dump of the reaction wheels to divide the LC in 24 chunks. Momentum dumps can produce jumps that could bias our analysis. We then fit a model composed of a mean shift for each chunk and a GP with a quasi-periodic kernel implemented using the \texttt{celerite} Python package \citep{foreman-mackey2017, foreman-mackey2018}. The functional form of the GP kernel is }
\begin{equation}\
\label{rotkernel}
    k(\tau) = \frac{B}{2 + C} e^{-\tau / L} \left[\cos\left(\frac{2 \pi\,\tau}{P_{\mathrm{rot}}}\right) + ( 1 + C)\right],
\end{equation}
{where $P_{\mathrm{rot}}$ is an estimator of the stellar rotation period, $L$ is the correlation timescale, $B$ is a positive amplitude term, and $C$ is a positive factor \citep[][eq. 56]{foreman-mackey2017}. We optimized the fit using a pre-optimization with a Nelder-Mead simplex algorithm \citep{NelderMead} using the Python package \texttt{scipy.optimize} followed by an MCMC exploration maximizing posterior probability of the model using \texttt{emcee}. We used multi-dimensional gaussian distribution for the likelihood. For the priors, we used log-uniform priors: between 0.1 ppm and 1 for $B$, between 1 and $10^4$ days for $L$,  between 1 and 100 days for $P_{\mathrm{rot}}$ and between $\exp(-5)$ and $\exp(5)$ for $C$. The \texttt{emcee} used 64 walker and a first exploration of 5\,000 iterations per walkers followed by a second exploration of 10\,000 iterations starting from the last position of the previous exploration. The result provides a broad and assymetric posterior distribution for the rotation period of $P_{\rm rot} = 30^{+38}_{-18}$ days. This result is then still inconclusive about the actual rotation period of the star, but certainly points to values in this regime.}

\begin{figure}
\centering
\includegraphics[width=0.5\textwidth]{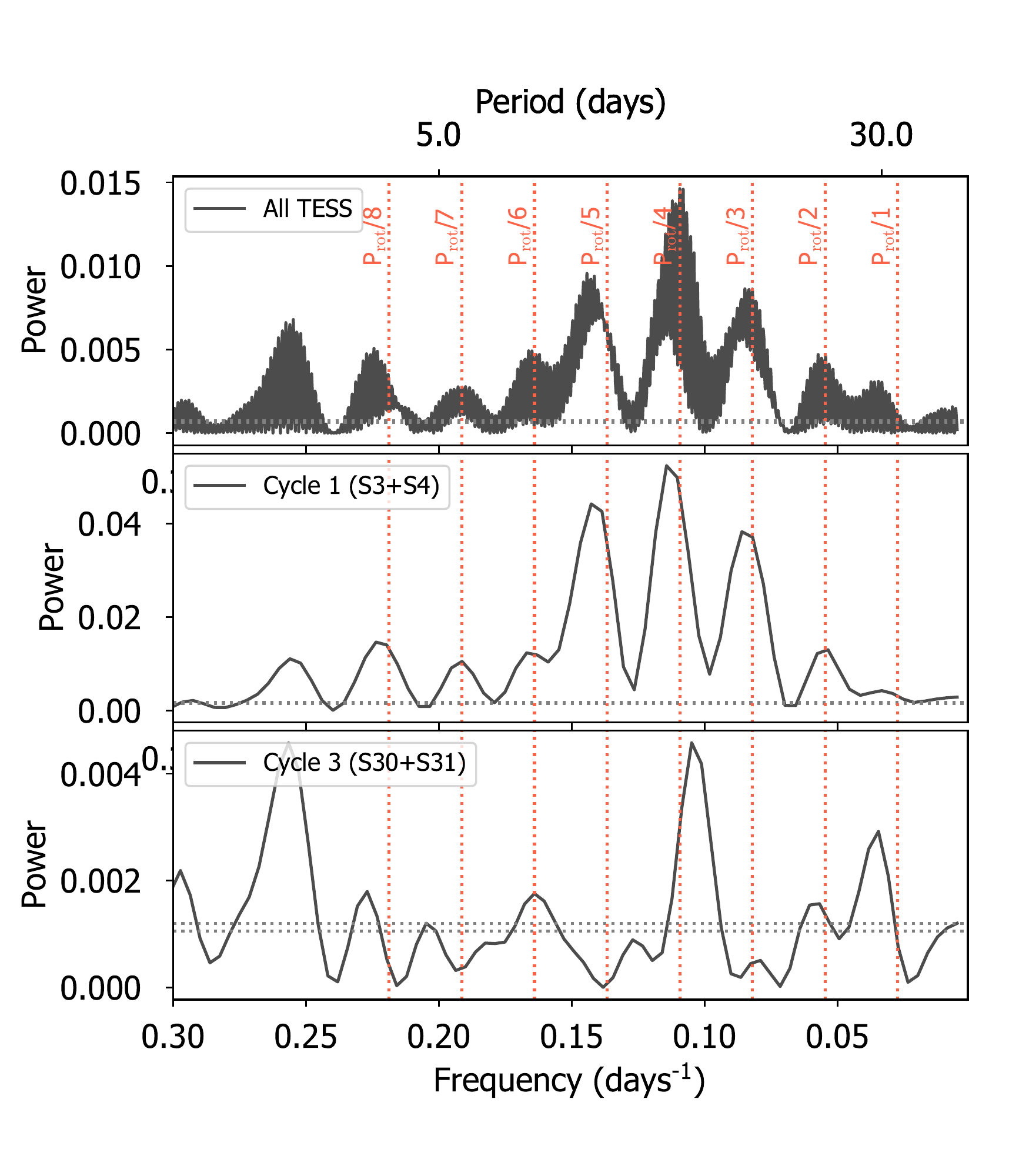}
\caption{{TESS light curve generalized Lomb-Scargle periodogram. Upper panel: GLSP using the full TESS dataset. Middle panel: GLSP using TESS data from Cycle 1, including sector 3 and 4. Lower panel: GLSP using TESS data from Cycle 3, including sector 30 and 31. In all panels, we highlight the first eight harmonics of the fundamental frequency corresponding to the expected rotation period of the star according to our RV analysis (see Sect.~\ref{sec:Analysis}) and from the expected value from the empirical relations by \cite{suarez-mascareno15}. }}
\label{fig:TESS_GLSP}
\end{figure}

{Finally, to further investigate the periodicities present in the LC, we computed its Generalized Lomb-Scargle periodogram (GLSP). However, to mitigate the impact of jumps between the different chunks of the LC, we first corrected the mean level of each chunks using mean shift obtained by the previous fit. The resulting periodogram is shown in Fig.~\ref{fig:TESS_GLSP}. Given the large difference in the amplitude of the LC variability between TESS cycle 1 (sectors 3 and 4) and TESS cycle 3 (sectors 30 and 31), we also include in this figure the individual GLSP of each cycle. While cycle 1 shows clear long-term variability, cycle 3 shows a clear decrease in the amplitude of the variations. We interpret this as different instances of the stellar magnetic cycle. However, we have no additional data to test this hypothesis. Interestingly, the full GLSP (driven by variations from cycle 1) displays strong peaks at the different harmonics of the fundamental frequency corresponding to the expected (from empirical relations by \citealt{suarez-mascareno15}) and measured (from our RV analysis in Sect.~\ref{sec:Analysis}) rotation period of around 35 days. However, the GLSP does not show any significant power at this fundamental frequency, while displaying a maximum power in the fourth harmonic (corresponding to a period of around 9 days). An in-depth explanation of this fact is out of the scope of this paper.} 

\section{Figures}

\section{Tables}
\newpage

\begin{table}[]
\setlength{\extrarowheight}{3pt}
\caption{\label{tab:basic} General properties of  \hd{}.}
\begin{tabular}{lll}
 
 \hline 
Parameter & Value & Ref.$^{\dagger}$ \\ 
\hline 
IDs & \hd{}, LHS\,1563,   \\
        &  GJ\,146, HIP\,16711 \\
Gaia EDR3 ID & 4833654227548585856  & [1] \\
RA, DEC & 03:35:01.79, -48:25:02.46 &  [1]\\ 
Parallax (mas) & $73.520\pm0.016$    & [1]\\
Distance (pc) & $13.602\pm0.003$ & [1, 3] \\
$\mu_{\alpha}$ (mas/yr) & $404.440\pm 0.018$  & [1]\\
$\mu_{\delta}$ (mas/yr) & $307.498\pm 0.023$  & [1]\\
RV (km/s) & $21.44 \pm 0.14$  &  [1] \\ 
G (mag) & 8.02  &  [1]\\
$B_p-R_p$ (mag) & 1.64  &  [1]\\
J (mag) & $6.122\pm0.020$ & [2] \\
Ks (mag) & $5.338\pm0.018$ & [2] \\
U (km/s) & -23.834 & [3], Sect.~\ref{sec:stellarprop}\\
V (km/s) & -7.957 & [3], Sect.~\ref{sec:stellarprop} \\
W (km/s) & 0.821 & [3], Sect.~\ref{sec:stellarprop}\\
Gal. population & Thin disk & [3], Sect.~\ref{sec:stellarprop} \\
T$_{\rm eff}$ (K) & $4385 \pm 21$ & [3], Sect.~\ref{sec:stellarprop} \\
$\log{\rm g}$ (dex)& $4.69\pm0.05$ & [3], Sect.~\ref{sec:stellarprop} \\
${\rm [Fe/H]}$ (dex) & $-0.08\pm 0.02$ & [3], Sect.~\ref{sec:stellarprop} \\
M$_{\star}$ (\Msun{}) & $0.684 \pm 0.013$ & [4] \\
R$_{\star}$ (\Rsun{}) & $0.674 \pm 0.020$ & [4] \\

 \hline

\end{tabular}
\tablebib{
[1] Gaia Collaboration et al. (2021); 
[2] Cohen et al. (2003); 
[3] This work;
[4] Mann et al. (2015).
}
\end{table}

\begin{table*}
\setlength{\extrarowheight}{3pt}
\caption{Radial velocities from HARPS and ESPRESSO used in this paper. The full table is accessible through CDS.}
\label{tab:RVespresso}
\begin{tabular}{lllll}
\hline\hline
BJD-2453000 (days) & RV (km/s) & FWHM (km/s) & BIS (km/s) & Instrument \\
\hline

-55.20791427 & $21.5911 \pm 0.0036$ & $6.1820 \pm 0.0022$ & $0.0340 \pm 0.0036$ & HARPS \\
668.81251209 & $21.5845 \pm 0.0015$ & $6.14227 \pm 0.00086$ & $0.0376 \pm 0.0015$ & HARPS \\
670.73193403 & $21.5755 \pm 0.0018$ & $6.1549 \pm 0.0011$ & $0.0365 \pm 0.0018$ & HARPS \\
721.71012278 & $21.5802 \pm 0.0017$ & $6.1508 \pm 0.0010$ & $0.0421 \pm 0.0017$ & HARPS \\
787.55098765 & $21.5808 \pm 0.0020$ & $6.1552 \pm 0.0012$ & $0.0332 \pm 0.0020$ & HARPS \\
... & & & & \\
2434.86710565 & $21.5842 \pm 0.0018$ & $6.1455 \pm 0.0010$ & $0.0383 \pm 0.0018$ & HARPS \\
2446.82486473 & $21.5824 \pm 0.0026$ & $6.1447 \pm 0.0016$ & $0.0491 \pm 0.0026$ & HARPS \\
2451.90757062 & $21.5803 \pm 0.0023$ & $6.1396 \pm 0.0014$ & $0.0355 \pm 0.0023$ & HARPS \\
2456.80553483 & $21.5768 \pm 0.0018$ & $6.1351 \pm 0.0010$ & $0.0306 \pm 0.0018$ & HARPS \\
2464.80400338 & $21.5830 \pm 0.0024$ & $6.1526 \pm 0.0015$ & $0.0483 \pm 0.0024$ & HARPS \\
5416.59771324 & $21.61391 \pm 0.00024$ & $6.49401 \pm 0.00048$ & $0.05932 \pm 0.00048$ & ESPRESSO18 \\
5421.70852604 & $21.61538 \pm 0.00016$ & $6.50414 \pm 0.00032$ & $0.05644 \pm 0.00032$ & ESPRESSO18 \\
5424.74838083 & $21.61737 \pm 0.00016$ & $6.50757 \pm 0.00032$ & $0.05648 \pm 0.00032$ & ESPRESSO18 \\
5444.74359584 & $21.61463 \pm 0.00017$ & $6.50204 \pm 0.00035$ & $0.05829 \pm 0.00035$ & ESPRESSO18 \\
5721.8415055 & $21.61126 \pm 0.00018$ & $6.52330 \pm 0.00035$ & $0.04938 \pm 0.00035$ & ESPRESSO19 \\
5725.85342698 & $21.61051 \pm 0.00019$ & $6.53435 \pm 0.00038$ & $0.05568 \pm 0.00038$ & ESPRESSO19 \\
5731.84188704 & $21.60716 \pm 0.00015$ & $6.52348 \pm 0.00030$ & $0.05413 \pm 0.00030$ & ESPRESSO19 \\
5737.84387896 & $21.61144 \pm 0.00017$ & $6.51981 \pm 0.00033$ & $0.05362 \pm 0.00033$ & ESPRESSO19 \\
5741.71207542 & $21.60762 \pm 0.00017$ & $6.52044 \pm 0.00034$ & $0.05494 \pm 0.00034$ & ESPRESSO19 \\
... & & & & \\
6210.6525173 & $21.60479 \pm 0.00015$ & $6.46166 \pm 0.00030$ & $0.05215 \pm 0.00030$ & ESPRESSO21 \\
6211.61306191 & $21.60626 \pm 0.00018$ & $6.46587 \pm 0.00037$ & $0.04840 \pm 0.00037$ & ESPRESSO21 \\
6212.62351337 & $21.60680 \pm 0.00019$ & $6.45961 \pm 0.00038$ & $0.04980 \pm 0.00038$ & ESPRESSO21 \\
6213.55749428 & $21.60550 \pm 0.00014$ & $6.46201 \pm 0.00029$ & $0.04990 \pm 0.00029$ & ESPRESSO21 \\
6214.57937965 & $21.60283 \pm 0.00032$ & $6.46251 \pm 0.00063$ & $0.05153 \pm 0.00063$ & ESPRESSO21 \\
... & & & & \\
\hline
\end{tabular}
\end{table*}

\begin{table*}
\setlength{\extrarowheight}{3pt}
\caption{Inferred and derived parameters for the one-planet model.}
\label{tab:posterior}
\begin{tabular}{lll}
\hline\hline
Parameter & Priors & Posteriors \\
\hline
\textit{Orbital parameters} & & ESPRESSO+HARPS \\
\hline
Orbital period, $P_b$ [days] & $\mathcal{G}$(5.09,1.0) & $5.09071^{+0.00026}_{-0.00026}$ \\
Time of inf. conjunction, $T_{\rm 0,b}-2400000$ [days] & $\mathcal{U}$(58600.0,58610.0) & $58602.560^{+0.050}_{-0.051}$ \\
RV semi-amplitude, $K_{\rm b}$ [m/s] & $\mathcal{U}$(0.0,100.0) & $2.62^{+0.23}_{-0.21}$ \\

\hline
\textit{Derived parameters} & & \\
\hline
Planet mass, $m_{b}\sin{i_b}$ [\Mearth{}] & (derived) & $5.57^{+0.73}_{-0.68}$ \\
Orbit semi-major axis, $a_{b}$ [AU] & (derived) & $0.0510^{+0.0024}_{-0.0026}$ \\
Relative orbital separation, $a_{b}/R_{\star}$ & (derived) & $24.3^{+7.1}_{-4.6}$ \\
Stellar effective incident flux, $S_{b}$ [$S_{\oplus}$] & (derived) & $26^{+13}_{-10}$ \\
Stellar luminosity, $L_{\star}$ [$L_{\odot}$] & (derived) & $0.067^{+0.033}_{-0.026}$ \\
Equilibrium temperature, $T_{\rm eq,b}$ [K] & (derived) & $573^{+63}_{-69}$ \\

\hline
\textit{GP parameters} & & \\
\hline
$\eta_{\rm 1,FWHM}$ [m/s] & $\mathcal{LU}$(0.01,400.0) & $17.0^{+2.8}_{-3.1}$ \\
$\eta_1$ [m/s] & $\mathcal{LU}$(0.01,150.0) & $2.94^{+0.58}_{-0.43}$ \\
$\eta_2$ [days] & $\mathcal{U}$(50.0,500.0) & $71.0^{+24}_{-9.3}$ \\
$\eta_3$ [days] & $\mathcal{U}$(20.0,50.0) & $34.99^{+0.58}_{-0.53}$ \\
$\eta_4$ & $\mathcal{LU}$(0.13,7.38) & $0.537^{+0.097}_{-0.079}$ \\

\hline
\textit{Instrument-dependent parameters} & & \\
\hline
$\delta_{\rm ESPRESSO18}$ [km/s] & $\mathcal{U}$(21.0,22.0) & $21.6164^{+0.0025}_{-0.0025}$ \\
$\delta_{\rm ESPRESSO19}$ [km/s] & $\mathcal{U}$(21.0,22.0) & $21.6088^{+0.0017}_{-0.0016}$ \\
$\delta_{\rm ESPRESSO21}$ [km/s] & $\mathcal{U}$(21.0,22.0) & $21.6085^{+0.0019}_{-0.0019}$ \\
$\delta_{\rm HARPS}$ [km/s] & $\mathcal{U}$(21.0,22.0) & $21.5815^{+0.0010}_{-0.0011}$ \\
$\sigma_{\rm ESPRESSO18}$ [m/s] & $\mathcal{LU}$(0.01,30.0) & $0.32^{+1.0}_{-0.29}$ \\
$\sigma_{\rm ESPRESSO19}$ [m/s] & $\mathcal{LU}$(0.01,30.0) & $0.16^{+0.53}_{-0.13}$ \\
$\sigma_{\rm ESPRESSO21}$ [m/s] & $\mathcal{LU}$(0.01,30.0) & $0.43^{+0.17}_{-0.13}$ \\
$\sigma_{\rm HARPS}$ [m/s] & $\mathcal{LU}$(0.1,30.0) & $3.01^{+0.63}_{-0.49}$ \\
$\delta_{\rm FWHM,ESPRESSO18}$ [km/s] & $\mathcal{G}$(6.5,0.1) & $6.503^{+0.014}_{-0.014}$ \\
$\delta_{\rm FWHM,ESPRESSO19}$ [km/s] & $\mathcal{G}$(6.5,0.1) & $6.5083^{+0.0088}_{-0.0094}$ \\
$\delta_{\rm FWHM,ESPRESSO21}$ [km/s] & $\mathcal{G}$(6.5,0.1) & $6.479^{+0.010}_{-0.011}$ \\
$\delta_{\rm FWHM,HARPS}$ [km/s] & $\mathcal{G}$(6.5,0.2) & $6.1500^{+0.0061}_{-0.0062}$ \\
$\sigma_{\rm FWHM,ESPRESSO18}$ [m/s] & $\mathcal{LU}$(0.1,100.0) & $1.4^{+11}_{-1.1}$ \\
$\sigma_{\rm FWHM,ESPRESSO19}$ [m/s] & $\mathcal{LU}$(0.1,100.0) & $1.7^{+2.3}_{-1.4}$ \\
$\sigma_{\rm FWHM,ESPRESSO21}$ [m/s] & $\mathcal{LU}$(0.1,100.0) & $2.01^{+0.61}_{-0.44}$ \\
$\sigma_{\rm FWHM,HARPS}$ [m/s] & $\mathcal{LU}$(0.1,100.0) & $19.9^{+3.9}_{-2.9}$ \\

\hline

\end{tabular}
\end{table*}

\end{document}